\documentclass[12pt]{iopart}
\usepackage{iopams}  
\usepackage{Citesort}
\usepackage{color}
\newcommand{\bra}{\left\langle}
\newcommand{\ket}{\right\rangle}
\newcommand{\pder}[2]{\frac{\partial #1}{\partial  #2}}
\newcommand{\der}[2]{\frac{d #1}{d  #2}}
\newcommand{\bv}[1]{{\boldsymbol #1}}

\begin{document}
\title[Derivation of 
fluctuating hydrodynamics
from underdamped Langevin equation]{
Derivation of 
the nonlinear fluctuating hydrodynamic equation 
from underdamped Langevin equation}

\author{Takenobu Nakamura$^1$ and Akira Yoshimori$^2$}
\address{$^1$ Research Institute of Computational Sciences (RICS), 
National Institute of Advanced Industrial Science and Technology (AIST),
1-1-1 Umezono, Tsukuba 305-8568, Japan}
\address{$^2$ Department of Physics, Kyushu University, 
Fukuoka 812-8581, Japan}
\ead{takenobu.nakamura@aist.go.jp}

\begin{abstract}
We derive the fluctuating  
hydrodynamic equation for the number and momentum densities
exactly from the underdamped Langevin equation. 
This derivation is an extension of  
the Kawasaki-Dean formula in underdamped case.
The steady state probability distribution of the number and momentum
 densities field can be expressed by the kinetic and potential energies.
In the massless limit, the obtained  
fluctuating hydrodynamic equation reduces to the Kawasaki-Dean  
equation.
Moreover, the derived equation corresponds to the field  
equation derived from the canonical equation when the friction coefficient  
is zero.
\end{abstract}


\section{Introduction}

Field equation is widely employed in the 
studies on colloidal or liquid dynamics.
In the study on colloidal dynamics, 
some researchers have applied averaged density
field dynamics, which is called the time-dependent density
functional method 
\cite{Marconi99,Penna03JCP,Penna03PRE,Dzubiella03,Archer04JCP,Marconi00,Kawa94}.
This method has also
been successfully employed to study various phenomena observed in the
field of liquid dynamics, such as
solvation
\cite{Bagchi88,Chandra88,Yoshimori96,Yoshimori98-1,Yoshimori98-2,Yoshimori04,Kawa94},
transport phenomena \cite{Araki95},
and slow relaxation in supercooled
liquids \cite{Fuchizaki98-1,Fuchizaki98-2}.
Besides the average density field dynamics,
other researchers have also
developed theoretical expressions describing momentum density
fields \cite{Yamaguchi,Yamaguchi07}.

As compared to the direct calculation of particle dynamics,
field description is more useful for theoretical studies.
This is because we can estimate many
physical parameters, such as transport coefficients
from the correlation functions of field variables.
Thus, by using the field description, many researchers have formulated
approximations for the estimation of the physical parameters.
For example, the mode-coupling theory, 
which is known as a useful tool for approximation of the transport coefficients,
has been formulated by using field description
\cite{SL,Sjogren,Kirkpatrick86-1,Kirkpatrick86-2}.

While the field description is useful
for theoretical calculations, 
its correspondence with the particle description is not clear.
Therefore, the derivation of the field description
from the particle description is the fundamental problem 
in the studies on colloidal and liquid dynamics.
When field variables are not averaged, 
Dean has derived the field equation from
the overdamped Langevin model \cite{Dean}.
In a colloidal system,
the time-dependent density functional method
can be applied for the derivation of field equations 
from the overdamped Langevin equation
by averaging the density field on the basis of
some assumptions 
\cite{Marconi99,Marconi00,Archer04JCP}.
Recently, a method has been developed to derive field equations 
from the Liouville equation describing liquid dynamics
by using the projection operator
method \cite{Yoshimori99,Yoshimori05}.
In most cases, the derivation of the field description from
the particle description requires some approximations.

Very few studied have been carried out on 
the derivation of the field description in the nonlinear and
underdamped cases. In these cases, the inertial effect has to be considered.
In liquid dynamics, linear generalized Langevin equations 
including momentum density have been derived
for the field variables of a homogeneous system  \cite{SL}.
Linear generalized linear Langevin equations for an inhomogeneous
system have also been developed \cite{Yamaguchi,Yamaguchi07}.
However, nonlinear equations in the filed description have not been derived.
Therefore, in underdamped cases, phenomenological models have often
been employed \cite{DasMazenko}.

In Ref. \cite{Dean}, 
the evolution equation
of the density field is derived 
from the overdamped Langevin equation 
representing the particles interacting via the pairwise potential.
The derived equation is called as the `Kawasaki-Dean formula'.
In Ref. \cite{Dean}, 
a closed evolution equation for the density field 
is exactly derived by using It\'{o}'s formula \cite{Gardiner},
while the evolution
 equation is approximately derived using
other field models.
The steady-state probability distribution 
of the density field for the overdamped Langevin model is represented by
the bare pairwise potential term and the entropy term.
In contrast to the overdamped case,
there are no exact derivations of the evolution equation 
for the field variables in the underdamped cases. 
Here, a question arises whether we can extend the Kawasaki-Dean formula
to the underdamped Langevin equation.
The exact derivation
of the closed evolution equation for field variables is main issue of
this paper.

In Sec. \ref{mainpart},
we derive the closed evolution equation for the number density 
and the momentum density field
using the underdamped Langevin model.
For a system without dissipation,
the closed evolution equation corresponds to the field equation 
for a Hamiltonian system.
In Sec. \ref{property},
we discuss the properties of the
derived evolution equation.
In the Sec. \ref{deri:Fokker-Planck}, 
we calculate the steady state probability distribution
functional of the evolution equation 
by using a functional Fokker-Planck equation.
In Sec. \ref{toDean}, we derive the Kawasaki-Dean equation
from the evolution equation derived in Sec. \ref{mainpart}
to check the consistency 
between our model and other models.
Sec. \ref{concludingremarks} presents the concluding remarks.

\section{Derivation of the nonlinear fluctuating hydrodynamic
equation from underdamped Langevin equation}\label{mainpart}

We study $N$ Brownian particles 
suspended in a three-dimensional solvent 
at temperature $T$.
The motion of the $i$-th Brownian particle
is represented by its position $\bv x_i$ and momentum $\bv p_i$,
where $i=1,2,...,N$ 
and $\bv x_i\in[0,L]\times[0,L]\times[0,L]$.
We express the $\alpha$-th component 
of $\bv x_i$ as $x_i^\alpha$,
where $\alpha=1,2,$ and 3. That is, 
$\bv x_i=(x_i^1,x_i^2,x_i^3)$.
The Brownian particles interact 
via the pairwise potential $V(\bv x)$.
Each Brownian particle has the same mass $m$.
The motion of the $i$-th Brownian particle is
described by the underdamped Langevin equation as
\begin{eqnarray}
\der{\bv x_i}{t}&=\frac{\bv p_i}{m}
\label{xdev},\\
\der{\bv p_i}{t}&=
-\pder{U(\{\bv x_j\}_{j=1}^N)}{\bv x_i}
-\frac{\gamma}{m}\bv p_i +\sqrt{\gamma T}\bv R_i(t),
\label{pdev}
\end{eqnarray}
where $U(\{\bv x_i\}_{i=1}^N)$ denotes 
the total potential energy defined as
\begin{eqnarray}
U(\{\bv x_i\}_{i=1}^N)
\equiv&
\frac{1}{2}\sum_{i=1}^N\sum_{j=1,j\neq i}^N V(\bv x_i-\bv x_j).
\end{eqnarray}
The coefficient $\gamma$ is the friction constant and 
$\bv R_i(t)$ is the zero-mean Gaussian white noise satisfying
\begin{eqnarray}
\bra R_i^\alpha(t)R_j^\beta(t')\ket
=2\delta_{ij}\delta_{\alpha\beta}\delta(t-t'),
\label{noise:und}
\end{eqnarray}
where $\bra \cdot \ket$ represents 
the average value of $\bv R_i(t)$.

First, as described in Ref. \cite{Dean},
we introduce the density field $\rho(\bv x,t)$ given as
\begin{eqnarray}
\rho(\bv x,t)\equiv
\sum_{i=1}^N\delta(\bv x-\bv x_i(t)).
\label{def:rho}
\end{eqnarray}
To obtain the closed evolution equation of the fields
in the underdamped system,
we also introduce the momentum density fields $\bv g(\bv x,t)$ defined as
\begin{eqnarray}
\bv g(\bv x,t)
\equiv\sum_{i=1}^N\bv p_i(t)\delta(\bv x-\bv x_i(t)).
\label{def:g}
\end{eqnarray}
One can exactly derive the closed evolution equation of
the number density field
and momentum density fields
defined by equations
(\ref{def:rho}) and (\ref{def:g}).

Using these definitions, we derive the evolution
equation  using It\'{o}'s formula.
Expanding the stochastic variable $\rho(\bv x,t)$ 
defined in equation (\ref{def:rho}),
we obtain the evolution equation for the number density as 
\begin{eqnarray}
\pder{\rho(\bv x,t)}{t}
=-\bv\nabla\cdot\left(\frac{\bv g(\bv x,t)}{m}\right),
\label{rhodev}
\end{eqnarray}
where we have used equations (\ref{xdev}) and (\ref{def:g}).
This equation represents the continuous equation for the density field.
Similarly,
using equations (\ref{pdev}) and (\ref{def:g}),
we obtain the evolution equation for the momentum density 
\begin{eqnarray}
\pder{g^\alpha(\bv x,t)}{t}
=&-\frac{\gamma}{m}g^\alpha(\bv x,t)
+\xi^\alpha(\bv x,t)\nonumber\\
&-\rho(\bv x,t)\int d\bv x
\pder{V(\bv x-\bv y)}{x^\alpha}\rho(\bv y,t)
-\pder{M^{\alpha\beta}(\bv x,t)}{x^\beta},
\label{gdevproto}
\end{eqnarray}
where we have used Einstein's summation convention 
whenever a subscript is repeated in a term.
Here, $\xi^\alpha(\bv x,t)$ and $M^{\alpha\beta}$ are defined as
\begin{eqnarray}
\xi^\alpha(\bv x,t)\equiv&
\sum_{i=1}^N
\sqrt{\gamma T}R_i^\alpha(t)\delta(\bv x-\bv x_i)\label{def:spacetimenoise},\\
M^{\alpha\beta}(\bv x,t)\equiv&
\sum_{i=1}^N\frac{p_i^\alpha(t)p_i^\beta(t)}{m}
\delta(\bv x-\bv x_i(t)).\label{Mdef}
\end{eqnarray}

Trivially, the average of $\bv \xi$ is zero
from equation (\ref{def:spacetimenoise}).
Further, $\bv \xi$ is a multiplicative 
noise: 
time correlation depends on the instantaneous density fields.
The noise term in 
equation (\ref{def:spacetimenoise}) 
is rewritten in the form
\begin{eqnarray}
\xi^\alpha(\bv x,t)
=\sqrt{\Gamma^{\alpha\beta}(\bv x,t)T}\zeta^\beta(\bv x,t)
\label{spacetimenoise2},
\end{eqnarray}
where $\Gamma^{\alpha\beta}(\bv x,t)$ is defined by
\begin{eqnarray}
\Gamma^{\alpha\beta}(\bv x,t)=\gamma\rho(\bv x,t)\delta_{\alpha\beta}
\label{Gamma}
\end{eqnarray}
and $\bv \zeta$ is the space-time Gaussian white noise
satisfying
\begin{eqnarray}
\bra\zeta^\alpha(\bv x,t)\zeta^\beta(\bv x',t')\ket
=2\delta_{\alpha\beta}\delta(\bv x-\bv x')\delta(t-t')
\label{spacetimenoise}.
\end{eqnarray}

To obtain the closed evolution equation, we have to
make the following assumption for the trajectory of 
the positions of a particle $\{\bv x_i(t)\}_{i=1}^N$:
\begin{eqnarray}
\delta(\bv x_i(t)-\bv x_j(t))=\delta(\bv x_i(t)-\bv x_j(t))\delta_{ij}
\label{deltaij},
\end{eqnarray}
where $\delta_{ij}$ is the Kronecker delta.
Note that our aim is to
construct a map 
from the trajectory of the position and momentum of the particles 
$\{\bv x_i(t),\bv p_i(t)\}_{i=1}^N$
to the trajectory of density and momentum density fields
$[\rho_t,\bv g_t]\equiv \{\rho(\bv x,t), \bv g(\bv x,t)\}_{\bv x}$.
Then, equation (\ref{deltaij}) is satisfied when
no two particles occupy 
the same position simultaneously in the mapping.
Such an assumption is valid if particles interact through 
via a repulsive pairwise potential
and a discretization of space, which is
discussed in \ref{justification}.

Equation (\ref{Mdef}) is formally rewritten in the form
\begin{eqnarray}
M^{\alpha\beta}(\bv x,t)
&=&\sum_{i=1}^N\frac{p_i^\alpha(t)p_i^\beta(t)}{m}
\delta(\bv x-\bv x_i(t))
\frac{\sum_{j=1}^N\delta(\bv x-\bv x_j(t))}
{\sum_{k=1}^N\delta(\bv x-\bv x_k(t))}\nonumber\\
&=&\frac{1}{m\rho(\bv x,t)}
\sum_{i=1}^N\sum_{j=1}^N
p_i^\alpha(t)p_i^\beta(t)\delta(\bv x-\bv x_i(t))\delta(\bv x-\bv x_j(t)).
\label{Mdef2}
\end{eqnarray}
The infinite form included in (\ref{Mdef2})
is explained in \ref{justification}.
In the second step, 
we have used the definition of
the density field given by the equation (\ref{def:rho}).
Then, by using equation (\ref{deltaij}),
$M^{\alpha\beta}(\bv x,t)$ is
represented only by $[\rho_t, \bv g_t]$
in the form
\begin{eqnarray}
M^{\alpha\beta}(\bv x,t)
&=&\frac{1}{m\rho(\bv x,t)}
\sum_{i=1}^Np_i^\alpha(t)\delta(\bv x-\bv x_i(t))
\sum_{j=1}^Np_j^\beta(t)\delta(\bv x-\bv x_j(t))
\nonumber\\
&=&\frac{g^\alpha(\bv x,t)g^\beta(\bv x,t)}{m\rho(\bv x,t)}
\label{Mdef3},
\end{eqnarray}
where we have used the definition of the momentum density given by
equation (\ref{def:g}) 
in the last step.

Finally, substituting equation (\ref{Mdef3}) into equation (\ref{gdevproto}),
we obtain the evolution equation for the momentum density as follows:
\begin{eqnarray}
\pder{g^\alpha(\bv x,t)}{t}
=&-\Gamma^{\alpha\beta}(\bv x, t)
\frac{\delta H_K[\rho,\bv g]}
{\delta g^\beta(\bv x,t)}
+\sqrt{\Gamma^{\alpha\beta}(\bv x,t)T}\zeta^\beta(\bv x,t)
\nonumber\\
&
-\rho(\bv x,t)\pder{ \ }{x^\alpha}
\left(
\frac{\delta H_V[\rho]}{\delta\rho(\bv x,t)}
\right)
-\pder{ \ }{x^\beta}
\left(
\frac{g^\alpha(\bv x,t)g^\beta(\bv x,t)}{m\rho(\bv x,t)}
\right),
\label{gdev}
\end{eqnarray}
where we use the abbreviation for the functional derivative as
\begin{eqnarray}
\frac{\delta H_V[\rho]}{\delta \rho(\bv x,t)}
&=&\left. 
\frac{\delta H_V[\varphi]}{\delta \varphi(\bv x)}
\right|_{\varphi(\bv x)=\rho(\bv x,t)},\nonumber\\
\frac{\delta H_K[\rho,\bv g]}{\delta g^\alpha(\bv x,t)}
&=&\left.\frac{\delta H_K[\varphi,\bv \psi]}
{\delta \psi^\alpha(\bv x)}
\right|_{
\varphi(\bv x)=\rho(\bv x,t),
\bv\psi(\bv x)=\bv g(\bv x,t)}.
\label{abbreviation}
\end{eqnarray}
This abbreviation is used hereinafter.
Here, $H_V[\varphi]$ and $H_K[\varphi,\bv \psi]$ 
are functionals
for the functions $\varphi(\bv x)$ and $\bv \psi(\bv x)$, respectively,
and are defined as
\begin{eqnarray}
H_V[\varphi]&\equiv&\frac{1}{2}
\int d\bv x \int d\bv y
V(\bv x-\bv y)(\varphi(\bv x)\varphi(\bv y)-\delta(\bv x-\bv y)
\varphi(\bv x))
\label{def:HV},\\
H_K[\varphi,\bv \psi]&\equiv&
\int d\bv x\frac{\bv \psi(\bv x)^2}{2m\varphi(\bv x)}\label{def:HK}.
\end{eqnarray}
Clearly, the functionals in equations (\ref{def:HV}) and (\ref{def:HK}) correspond to 
the internal energy and the kinetic energy of the system, respectively.
Equations (\ref{rhodev}) and (\ref{gdev}) are the desired
 nonlinear fluctuating hydrodynamic equations.

The first term on the right-hand side of equation (\ref{gdev}) 
leads to the decay of momentum.
The dissipative matrix 
$\Gamma^{\alpha\beta}$ in equation (\ref{gdev}) depends on $\rho(\bv x)$.
This feature, which is a characteristic of the 
Brownian particle system, is in contrast to features of 
the Navier-Stokes equation.
The dissipative matrix in the Navier-Stokes equation is
given
by the combination of the gradient and the shear and bulk viscosities 
\cite{DasMazenko,Kirkpatrick86-1}.
The fluctuation-dissipation relation of
 the second kind is satisfied by the first and second terms
on the right-hand side of equation (\ref{gdev}).
That is, the dissipative matrix is consistent with the noise 
coefficient.

The third and fourth terms on the right-hand side of equation (\ref{gdev})
represent
the conservative flows. The flow represented by the fourth term
is caused by the momentum transfer.
Further, the flow represented by the third term is 
caused by the gradient of 
the functional derivative of the Hamiltonian 
including the bare potential $V(\bv x-\bv y)$ (or $H_V[\rho]$)
in equation (\ref{def:HV}).
This is in contrast to many field models including
the chemical potential or free energy.
The bare potential is obtained by the exact derivation
from the overdamped Langevin model \cite{Dean}.
Thus, the present result shows that the inclusion of the bare potential
is general consequence of the exact derivation 
without any coarse graining.

The Hamiltonian in equation (\ref{gdev}) does not include the entropy terms,
which are included in 
the overdamped evolution equation for the fields \cite{Dean} 
or in the phenomenological model in the underdamped case
\cite{Kawa94,Nishino}.
One can drive the entropy terms for the Brownian particle system 
from the momentum transfer term when the overdamped limit 
is considered in equation (\ref{gdev}) (Sec. \ref{toDean}).
The entropy terms for liquid dynamics also
originate from the momentum transfer term in the Liouville  
equation \cite{Yoshimori05}.
Those results indicates that 
the entropy term in the evolution equations
is eliminated by explicitly treating the  
momentum transfer term 
from the point of view of the derivation from a microscopic model.

From equations (\ref{rhodev}) and (\ref{gdev}),
we also obtain the closed evolution equation of the density
 and momentum density from the canonical equation.
Equations (\ref{xdev}) and (\ref{pdev}) reduced
 to the canonical equation when $\gamma =0$.
Therefore, by substituting $\gamma=0$ into 
equations (\ref{rhodev}) and (\ref{gdev}),
 we obtain
\begin{eqnarray}
\pder{\rho(\bv x,t)}{t}=&-\bv \nabla\cdot
\left[\frac{\bv g(\bv x,t)}{m}\right]
\label{Newton:field1},\\
\pder{\bv g(\bv x,t)}{t}=&-\rho(\bv x,t)\bv\nabla
\left[\frac{\delta H_V[\rho]}{\delta\rho(\bv x,t)}\right]
-\bv\nabla\cdot\left[
\frac{\bv g(\bv x,t)\bv g(\bv x,t)}{m\rho(\bv x,t)}
\right].
\label{Newton:field2}
\end{eqnarray}
The evolution equations 
(\ref{Newton:field1}) and (\ref{Newton:field2})
contain the following five conserved quantities:
the total energy, the total number, and total momentums.
These quantities are defined as
\begin{eqnarray}
H[\rho,\bv g]&\equiv H_V[\rho]+H_K[\rho,\bv g],\\
N[\rho]&\equiv \int d\bv x\rho(\bv x),\\
\bv P[\bv g]&\equiv \int d\bv x\bv g(\bv x).
\end{eqnarray}

The conservation law for the total energy functional
is derived as follows:
\begin{eqnarray}
\der{H[\rho_t,\bv g_t]}{t}
=&\int d\bv x\pder{\rho(\bv x,t )}{t}
\left[
\frac{\delta H_V[\rho]}{\delta \rho(\bv x,t)}
-\frac{g^2(\bv x,t)}{2m\rho^2(\bv x,t)}\right]\nonumber\\
&+\int d\bv x\pder{g^\alpha(\bv x,t )}{t}
\frac{g^\alpha(\bv x,t )}{m\rho(\bv x, t)}.\label{energycons0}
\end{eqnarray}
By substituting equations (\ref{Newton:field1}) and (\ref{Newton:field2})
into equation (\ref{energycons0})
and integrating by parts several times,
we have
\begin{eqnarray}
\der{H[\rho_t,\bv g_t]}{t}
&=-\int d\bv x\bv\nabla\cdot
\left[\frac{\bv g(\bv x, t)}{m}
\left(
\frac{\delta H_V[\rho]}{\delta \rho(\bv x, t)}
-\frac{\delta H_K[\rho,\bv g]}{\delta \rho(\bv x, t)}
\right)\right].\label{energycons2}
\end{eqnarray}
The left-hand side of 
equation (\ref{energycons2})
is equal to zero from the divergence theorem.
In addition, the conservation law for the total number
 can be easily checked from 
equation (\ref{Newton:field1}).

The conservation law for the total momentum is also proved as follows:
\begin{eqnarray}
\der{\bv P[\bv g_t]}{t}
=&-\int d\bv x\rho(\bv x,t)\bv\nabla\left[
\frac{\delta H_V[\rho]}{\delta \rho(\bv x,t)}\right]
-\int d\bv x\bv\nabla\cdot
\left[\frac{\bv g(\bv x,t)\bv g(\bv x,t)}{m\rho(\bv x,t)}\right].
\label{totalmomentum}
\end{eqnarray}
The first term on the right-hand side of 
equation (\ref{totalmomentum})
vanishes by the action-reaction law.
The second term on the 
right-hand side of 
equation (\ref{totalmomentum})
vanishes from  the divergence theorem.
Therefore, the total momentums are conserved.
Note that we have obtained the conservation law
directly from the continuous model given 
by equations (\ref{Newton:field1}) and (\ref{Newton:field2})
without using the canonical equations
 (\ref{xdev}) and (\ref{pdev}).

Equations (\ref{Newton:field1}) and (\ref{Newton:field2})
are similar to the Euler equation
 in fluid mechanics \cite{Landau}.
In these equations, the number and the momentum are conserved,
and the advection term is present.
However, there are some differences between them,
which will be discussed
in Sec. \ref{concludingremarks}.

\section{Properties of 
the nonlinear fluctuating hydrodynamic equation 
} \label{property}
In this section, we discuss some aspects of 
the closed stochastic evolution equations
(\ref{rhodev}) and (\ref{gdev}) 
along with the Hamiltonians (\ref{def:HV})
and (\ref{def:HK}) and
the noise given by equation (\ref{spacetimenoise}).

\subsection{Derivation of Fokker-Planck equation 
for underdamped fluctuating hydrodynamic equation}\label{deri:Fokker-Planck}

In this subsection,
we calculate the steady-state probability distribution 
functional for the number and momentum density fields
from the derived stochastic evolution equations 
(\ref{rhodev}) and (\ref{gdev}).
We first derive
 the Fokker-Planck equation for these field variables
by using a standard procedure.
Then, we obtain the 
steady-state probability distribution functional
as a stationary solution for the Fokker-Planck equation.
In this subsection, 
a time-dependent function
$f(\bv x,t)$ is denoted by $f_t(\bv x)$
using standard notations for a stochastic process.

The probability density distribution functional is defined as
\begin{eqnarray}
P([\rho,\bv g],t)=\bra\delta[\rho-\rho_t]\delta[\bv g- \bv g_t]\ket
\label{def:P},
\end{eqnarray}
where $\bra \cdot \ket$ represents the average
over $\{\bv \zeta_t(\bv x)\}_{\bv x}$.
$\delta[\cdot]$ is a delta functional defined as
\begin{eqnarray}
\delta[\rho-\rho_t]\delta[\bv g-\bv g_t]
\equiv \prod_{\bv x}\delta\left(\rho(\bv x)-\rho_t(\bv x)\right)
\delta\left(\bv g(\bv x)-\bv g_t(\bv x)\right).
\label{Phat}
\end{eqnarray}

The evolution equation
for the density field given by (\ref{rhodev})
is rewritten in the form
\begin{eqnarray}
d\rho_t(\bv x)=-\bv\nabla\cdot\bv g_t(\bv x)dt/m .
\label{rhodev:ito}
\end{eqnarray}
In addition, the evolution equations
for the momentum density fields given by (\ref{gdev})
is rewritten in the form
\begin{eqnarray}
d\bv g_t(\bv x)
&=\bv G(\rho_t(\bv x),\bv g_t(\bv x))dt
+\sqrt{\gamma \rho_t(\bv x)T}d\bv\eta_t(\bv x)
\label{gdev:ito}.
\end{eqnarray}
Here, $\bv G$ is defined as 
\begin{eqnarray}
\fl G^\alpha(\rho(\bv x),\bv g(\bv x))\equiv&
-\gamma\rho(\bv x)
\frac{\delta H_K[\rho,\bv g]}{\delta g^\alpha(\bv  x)}
-\rho(\bv x)\frac{ \partial }{\partial x^\alpha}
\left[\frac{\delta H_V[\rho]}{\delta\rho(\bv x)}\right]
-\frac{\partial}{\partial x^\beta}
\left[\frac{g^\alpha(\bv x)g^\beta(\bv x)}{m\rho(\bv x)}\right]
\label{def:G}
\end{eqnarray}
and $\eta^\alpha_t(\bv x)$ satisfies
\begin{eqnarray}
d\eta^\alpha_t(\bv x)d\eta^\beta_t(\bv x')
=2\delta_{\alpha\beta}\delta(\bv x-\bv x')dt.
\label{ito2}
\end{eqnarray}
Here, $d\eta_t^\alpha(\bv x)d\eta_{t'}^\beta(\bv x')$ is 
equal to zero in the case of  $t\neq t'$.

To obtain the Fokker-Planck equation, 
we apply It\'o's formula in (\ref{Phat}) as follows:
\begin{eqnarray}
\fl &&
d\left\{\delta[\rho-\rho_t]\delta[\bv g-\bv g_t]\right\}\nonumber\\
\fl&=&\int d\bv x d\rho_t(\bv x)
\frac{\delta \left\{\delta[\rho-\rho_t]\delta[\bv g-\bv g_t]\right\}}{\delta \rho_t(\bv x)}\nonumber\\
\fl&&+\int d\bv xd\bv g_t(\bv x)\cdot\frac{
\delta\left\{\delta[\rho-\rho_t]\delta[\bv g-\bv g_t]\right\}}{\delta \bv g_t(\bv x)}\nonumber\\
\fl&&+\frac{1}{2}\int d\bv xd\bv g_t(\bv x)\cdot
\frac{ \delta }{\delta \bv g_t(\bv x)}
\left[
\int d\bv x'
d\bv g_t(\bv x')
\cdot\frac{\delta \left\{\delta[\rho-\rho_t]\delta[\bv g-\bv g_t]\right\}}
{\delta \bv g_t(\bv x')}\right].
\label{hat:P}
\end{eqnarray}
Here, we have defined
the functional derivative as
\begin{eqnarray}
\fl \int d\bv x d\rho_t(\bv x)
\frac{ \delta 
\left\{\delta[\rho-\rho_t]\delta[\bv g-\bv g_t]\right\}}
{\delta \rho_t(\bv x)}
\nonumber\\
\fl \equiv \lim_{|\Delta \bv x|\to 0}\sum_{\bv I}
d\rho_t(\bv x_{\bv I})\pder{ \ }{\rho_t(\bv x_{\bv I})}
\prod_{\bv I'}\delta(\rho_t(\bv x_{\bv I'})-\rho(\bv x_{\bv I'}))
\delta(\bv g_t(\bv x_{\bv I'})-\bv g(\bv x_{\bv I'})),
\end{eqnarray}
where $\bv I$ and $\bv I'$
are the indices
of the discretized space coordinate
with volume $|\Delta \bv x|$
and $\bv x_{\bv I}$ is the discretized position.
By substituting the evolution equations (\ref{rhodev:ito}) and (\ref{gdev:ito})
into equation (\ref{hat:P}), we obtain
\begin{eqnarray}
\fl &
d\left\{\delta[\rho-\rho_t]\delta[\bv g-\bv g_t]\right\}
\nonumber\\
\fl=&
-\int d\bv x
\bv\nabla\cdot\bv g_t(\bv x)dt/m
\frac{\delta \left\{\delta[\rho-\rho_t]\delta[\bv g-\bv g_t]\right\}}
{\delta \rho_t(\bv x)}\nonumber\\
\fl &+\int d\bv x
\left(\bv G(\rho_t(\bv x),\bv g_t(\bv x))dt+
\sqrt{\gamma T\rho_t(\bv x)}d\bv\eta_t(\bv x)\right)
\cdot\frac{\delta 
\left\{\delta[\rho-\rho_t]\delta[\bv g-\bv g_t]\right\}
}{\delta \bv g_t(\bv x)}\nonumber\\
\fl &+\frac{1}{2}\int d\bv x
\sqrt{\gamma T\rho_t(\bv x)}d\bv\eta_t(\bv x)\nonumber\\
\fl &\cdot
\frac{ \delta  }{\delta \bv g_t(\bv x)}
\left[\int d\bv x'
\sqrt{\gamma T\rho_t(\bv x')}d\bv\eta_t(\bv x')
\cdot
\frac{\delta \left\{\delta[\rho-\rho_t]\delta[\bv g-\bv g_t]\right\}}
{\delta \bv g_t(\bv x')}\right]\label{deriFP1}.
\end{eqnarray}
Changing the index of the derivative of the
 delta functional from $[\rho_t,\bv g_t]$ to $[\rho,\bv g]$,
 we obtain
\begin{eqnarray}
\fl &d\left\{\delta[\rho-\rho_t]\delta[\bv g-\bv g_t]\right\}\nonumber\\
\fl =&\int d\bv x
\frac{ \delta }{\delta\rho(\bv x)}\left[
\bv\nabla\cdot\bv g(\bv x)dt/m
\left\{\delta[\rho-\rho_t]\delta[\bv g-\bv g_t]\right\}\right]\nonumber\\
\fl &-\int d\bv x
\frac{ \delta }{\delta \bv g(\bv x)}
\cdot\left[
\left(\bv G(\rho(\bv x),\bv g(\bv x))dt
+\sqrt{\gamma T\rho_t(\bv x)}d\bv\eta_t(\bv x)\right)
\left\{\delta[\rho-\rho_t]\delta[\bv g-\bv g_t]\right\}\right]\nonumber\\
\fl &+\int d\bv x
\frac{ \delta }{\delta \bv g(\bv x)}\cdot
\frac{ \delta }{\delta \bv g(\bv x)}\left[
\gamma T\rho(\bv x)dt
\left\{\delta[\rho-\rho_t]\delta[\bv g-\bv g_t]\right\}
\right]\label{nonaverage:result}.
\end{eqnarray}
Here, we have integrated the third
term on the right-hand side of equation (\ref{nonaverage:result})
 with respect to $\bv x'$ using equation (\ref{ito2}).
By substituting equation (\ref{def:G}) into equation (\ref{nonaverage:result})
and averaging (\ref{nonaverage:result}),
we obtain
the Fokker-Planck equation for $[\rho, \bv g]$ as
\begin{eqnarray}
\pder{P([\rho,\bv g],t)}{t}
=\hat {\cal L}([\rho,\bv g])P([\rho,\bv g],t)\label{Fokker-Planck:und}.
\end{eqnarray}
Here, the operator $\hat {\cal L}$ is a linear operator 
defined as 
\begin{eqnarray}
\hat {\cal L}([\rho,\bv g])
\equiv&\int d\bv x\left\{
\frac{\delta}{\delta\rho(\bv x)}
\bv\nabla\cdot
\left(\frac{\bv g(\bv x)}{m}\right)\right.
\nonumber\\
&+\frac{\delta}{\delta\bv g(\bv x)}
\cdot\left[
\rho(\bv x)\bv \nabla
\left(\frac{\delta H_V[\rho]}{\delta\rho(\bv x)}\right)
+
\bv\nabla\cdot\left(
\frac{\bv g(\bv x)\bv g(\bv x)}
{m\rho(\bv x)}\right)
\right]\nonumber\\
&+\left.
\frac{\delta}{\delta\bv g(\bv x)}
\cdot{\bf \Gamma}(\bv x)\cdot
\left(T\frac{\delta}{\delta\bv g(\bv x)}
+\frac{\delta H_K[\rho,\bv g]}{\delta \bv g(\bv x)}
\right)
\right\}\label{operatorL},
\end{eqnarray}
where 
$\Gamma^{\alpha\beta}(\bv x)=\gamma\rho(\bv x)\delta_{\alpha\beta}$.
We obtain the steady-state probability distribution functional 
$P_{\rm eq}[\rho,\bv g]$
as a stationary solution of 
equation (\ref{Fokker-Planck:und})
given by
\begin{eqnarray}
P_{\rm eq}[\rho,\bv g]
=\frac{1}{Z}\exp\left(
-\frac{H_V[\rho]+H_K[\rho,\bv g]}{T}\right)
\label{steadystatedistribution:und},
\end{eqnarray}
where $Z$ is a normalization constant determined by 
\begin{eqnarray}
\int {\cal D}\rho{\cal D}\bv g
P_{\rm eq}[\rho,\bv g]=1.\label{normalization}
\end{eqnarray}

Note that $P_{\rm eq}[\rho,\bv g]$ does not include the entropy terms
although the Hamiltonian in
the steady-state distribution obtained using the overdamped field model
includes them \cite{Dean}.
Further, the steady-state distribution functional
$P_{\rm eq} [\rho,\bv g]$ has same form as
the classical fluid \cite{Kawa94}.

\subsection{Massless limit of 
the underdamped fluctuating hydrodynamics}\label{toDean}
In this subsection,
 from equations (\ref{rhodev}) and (\ref{gdev}),
we derive the overdamped fluctuating hydrodynamic equation 
for Brownian particle systems 
in the massless limit.
In the massless limit, the obtained equation
 is the so-called Kawasaki-Dean formula.
Note that 
the Kawasaki-Dean formula is derived 
from the overdamped Langevin equation,
which is obtained from the underdamped Langevin equation 
in the massless limit.
Therefore, the derivation of Kawasaki-Dean formula from 
our equations leads to a consistency between our equations 
and these equations. 
Moreover, the derivation given in this section is useful
for understanding similar studies carried out
in the past \cite{Kawa94,Yoshimori05}.

Using equations (\ref{def:rho}) and (\ref{deltaij}), we obtain the identity
\begin{eqnarray}
\rho(\bv x,t)\rho(\bv x,t)=\delta(\bv x-\bv x)\rho(\bv x,t).
\label{identity:rhorho}
\end{eqnarray}
Here, the right-hand side of the identity has an infinite value,
which can be 
justified by the proper interpretation of discretization 
discussed
in \ref{justification}.
Equation (\ref{identity:rhorho}) is satisfied 
only if the density is defined as the sum of delta functions.
Therefore if the density is defined as a continuous function,
equation (\ref{identity:rhorho}) is not satisfied.

By using $\tau\equiv m/\gamma$, 
the evolution equations of the density field and 
momentum density field are rewritten in the form
\begin{eqnarray}
\pder{\rho(\bv x,t)}{t}&=&-\frac{1}{\tau\gamma}
 \bv \nabla\cdot \bv g(\bv x,t)\label{rhoev:pre2},\\
\pder{g^\alpha(\bv x,t)}{t}
&=&
-\frac{g^\alpha(\bv x,t)}{\tau}
-\rho(\bv x,t)\frac{\partial}{\partial x^\alpha}\left[
\frac{\delta H_V[\rho]}{\delta\rho(\bv x,t)}\right]
 \nonumber\\
&&-\frac{\partial}{\partial x^\beta}
\left[\frac{g^\alpha(\bv x,t)g^\beta(\bv x,t)}{\tau\gamma\rho(\bv x,t)}\right]
 +\sqrt{\gamma T\rho(\bv x,t)}\zeta^\alpha(\bv x,t)
\label{gev:pre2}.
\end{eqnarray}

The parameter $\tau$ is the relaxation time for the density field
and is constant for a given system.
We focus on the time evolution of the density field
whose time resolution $\Delta t$ is significantly larger 
than $\tau$.
Then, we define
\begin{eqnarray}
\tilde\rho(\bv x,t_n)\equiv 
\lim_{\tau/\Delta t \to 0}\rho(\bv x,t_n),
\label{def:tilderho}
\end{eqnarray}
where $t_n=n\Delta t$.
After taking the limit of $\tau$ and evaluating the equation,
we take the continuous limit $\Delta t\to 0$
and represent the time evolution 
of the coarse-grained density field $\tilde\rho$ as follows:
\begin{eqnarray}
\pder{\tilde\rho(\bv x,t)}{t}
&\equiv& \lim_{\Delta t\to 0}
\left[\frac{\tilde\rho(\bv x,t_n+\Delta t)-\tilde\rho(\bv x,t_n)}
{\Delta t}\right]\nonumber\\
&=&\lim_{\Delta t\to 0}
\left[\lim_{\tau\to 0}
\left[
\frac{\rho(\bv x,t+\Delta t)-\rho(\bv x,t)}{\Delta t}\right]
\right]\label{def:evolvetilderho}.
\end{eqnarray}

In the derivation of the coarse-grained evolution equation,
we have used the asymptotic formula
\begin{eqnarray}
\lim_{\tau\to 0}\int^{t'}_{t'_-} dt
\frac{e^{-(t'-t)/\tau}}{\tau}A(t)=A(t')
{\rm \ \ for \ \ }t'>t'_-.
\label{formula:deltaintegral}
\end{eqnarray}
That is because a term in the integrand is used in the definition
of the delta function
\begin{eqnarray}
\delta(t-t')=\lim_{\tau\to 0}
\frac{e^{-|t-t'|/\tau}}{2\tau}.\label{formula:delta}
\end{eqnarray}
Note that $t'$ is the upper limit of the integral
in equation (\ref{formula:deltaintegral}).

By integrating (\ref{rhoev:pre2}) with respect to time,
we obtain the difference $\rho(\bv x,t+\Delta t)-\rho(\bv x,t)$ in 
equation (\ref{def:evolvetilderho}) as follows:
\begin{eqnarray}
\rho(\bv x,t+\Delta t)-\rho(\bv x,t)
=-\frac{1}{\gamma}
\bv{\nabla}\cdot\int_t^{t+\Delta t}dt'\frac{1}{\tau}\bv g(\bv x,t').
\label{integrate:rho}
\end{eqnarray}
Next, we consider a system with $t \gg \tau$.
By using equation (\ref{gev:pre2}), 
$\bv g(\bv x,t')$ is formally solved as
\begin{eqnarray}
\frac{1}{\tau}\bv g(\bv x,t')=\frac{1}{\tau}\bv g(\bv x,0)e^{-t'/\tau}
+\bv\Upsilon(\bv x,t')+\bv\Xi(\bv x,t')+\bv\Pi(\bv x,t').
\label{solution:g}
\end{eqnarray}
Here, we have introduced the quantities
\begin{eqnarray}
\Upsilon^\alpha(\bv x,t')
&\equiv
-\frac{1}{\tau}\int_0^{t'} ds e^{-(t'-s)/\tau}
\rho(\bv x,s)\frac{\partial}{\partial x^\alpha}
\left[
\frac{\delta H_V[\rho]}{\delta\rho(\bv x,s)}\right]
\label{def:Upsilon},\\
\Xi^\alpha(\bv x,t')
&\equiv\frac{1}{\tau}\int_{s=0}^{s=t'} 
e^{-(t'-s)/\tau}
\sqrt{\gamma T\rho(\bv x,s)}d\eta^\alpha_s(\bv x)
\label{def:Xi},\\
\Pi^\alpha(\bv x,t')
&\equiv-\frac{1}{\tau}\int_0^{t'} ds e^{-(t'-s)/\tau}
\frac{\partial}{\partial x^\beta}
\left[\frac{g^\alpha(\bv x,s)g^\beta(\bv x,s)}
{\tau \gamma\rho(\bv x,s)}\right]
\label{def:Pi}.
\end{eqnarray}
The terms
$\bv\Xi$, $\bv\Upsilon$, and $\bv\Pi$ correspond to
noise, drift, and advection terms respectively.
Substituting
equation (\ref{solution:g})
into equation (\ref{integrate:rho}), we obtain
\begin{eqnarray}
\fl \gamma\left(\rho(\bv x,t+\Delta t)-\rho(\bv x,t)\right)
=-\bv\nabla\cdot\left[
\int_t^{t+\Delta t}dt'
\left(\bv\Upsilon(\bv x,t')+\bv \Xi(\bv x,t')+\bv\Pi(\bv x,t')\right)
\right].
\label{rho:base}
\end{eqnarray}

Using equations
(\ref{def:Upsilon}), (\ref{def:Xi}), (\ref{def:Pi}), 
and (\ref{rho:base}),
we evaluate the right-hand side of 
equation (\ref{def:evolvetilderho})
as follows.
First, we substitute
equations (\ref{def:Upsilon}), (\ref{def:Xi}),
and (\ref{def:Pi}) recursively into the left-hand side of
equation (\ref{rho:base}).
Next, taking the limit $\tau\to 0$,
we evaluate it
to the first order of $\Delta t$.
Then, taking the continuous limit $\Delta t\to 0$,
we obtain the right-hand side of 
equation (\ref{def:evolvetilderho}).

First, we integrate of $\Upsilon$.
By using 
equation (\ref{formula:deltaintegral}),
equation (\ref{def:Upsilon}) is evaluated as 
\begin{eqnarray}
\lim_{\tau\to 0}\Upsilon^\alpha(\bv x,t')
=-\tilde\rho(\bv x,t')\frac{\partial}{\partial x^\alpha}
\left[\frac{\delta H_V[\tilde\rho]}{\delta \tilde\rho(\bv x ,t')}\right].
\end{eqnarray}
Then, the integral of $\Upsilon(\bv x,t')$ in the limit $\tau\to 0$
is evaluated as
\begin{eqnarray}
\lim_{\tau\to 0}\int_t^{t+\Delta t}dt'\Upsilon^\alpha(\bv x,t')
=&-\tilde\rho(\bv x,t)\frac{\partial}{\partial x^\alpha}\left[
\frac{\delta H_V[\tilde\rho]}{\delta\tilde\rho(\bv x,t)}\right]\Delta t.
\label{result:Upsilon}
\end{eqnarray}

Next, to integrate $\bv\Xi$,
we calculate the correlation for these variables 
in the case of $\Delta t\gg \tau$.
The product $\Xi^\alpha(\bv x,t_1)\Xi^\beta(\bv x',t_2)$ is 
calculated as
\begin{eqnarray}
\fl \Xi^\alpha(x,t_1)\Xi^\beta(x',t_2)
&=&\int_{s_1=0}^{s_1=t_1}\int_{s_2=0}^{s_2=t_2}
\frac{e^{-(t_1+t_2-s_1-s_2)/\tau}}{\tau^2}\gamma T
\sqrt{\rho(\bv x,s_1)\rho(\bv x',s_2)}d\eta^{\alpha}_{s_1}(\bv x)
d\eta^{\beta}_{s_2}(\bv x')\nonumber\\
\fl &=&2\gamma T\delta(\bv x-\bv x')\delta^{\alpha\beta}e^{-|t_1-t_2|/\tau}
\int_{0}^{\min [t_1,t_2]}ds
\frac{e^{-2(\min[t_1,t_2]-s)/\tau}}{\tau^2}\rho(\bv x,s)
\label{xixim1}.
\end{eqnarray}
Here, we have used It\'{o} calculus (\ref{ito2})
and the identity $t_1 + t_2=|t_1 - t_2| + 2\min[t_1,t_2]$.
By integrating equation (\ref{xixim1}) with respect to time $t_1$ and $t_2$, 
we represent the product of integrations of $\Xi$ as
\begin{eqnarray}
\fl \int_t^{t+\Delta t}dt_1\int_t^{t+\Delta t}dt_2\Xi^\alpha(\bv x,t_1)
\Xi^\beta(\bv x',t_2)
=&2\gamma T\delta_{\alpha\beta}\delta(\bv x-\bv x')
\int_t^{t+\Delta t} dt_1\int_t^{t+\Delta t}
dt_2\frac{e^{-|t_1-t_2|/\tau}}{\tau}\nonumber\\
\fl &\times \int_0^{\min[t_1,t_2]} ds 
\frac{e^{-2(\min[t_1,t_2]-s)/\tau}}{\tau}
\rho(\bv x,s).
\label{xixi}
\end{eqnarray}
Taking the limit $\tau\to 0$ in 
equation (\ref{xixi}), we obtain
\begin{eqnarray}
\fl \lim_{\tau\to 0}
\int_t^{t+\Delta t}dt_1\int_t^{t+\Delta t}dt_2\Xi^\alpha(\bv x,t_1)
\Xi^\beta(\bv x',t_2)
=2\gamma T\delta_{\alpha\beta}\delta(\bv x-\bv x')\tilde\rho(\bv x,t)\Delta t
+o(\Delta t).
\label{XiXi}
\end{eqnarray}
Comparing 
equation (\ref{XiXi}) with 
equations (\ref{spacetimenoise2}), (\ref{Gamma}),
and (\ref{spacetimenoise}),
we find that 
the time average of $\bv\Xi(\bv x, t)$ coincides 
with that of $\bv\xi(\bv x,t)$ 
when $\rho$ is replaced with $\tilde\rho$.

Next, we integrate of $\Pi$.
By substituting equation (\ref{solution:g}) 
into equation (\ref{def:Pi}) recursively,
we integrate of $\Pi(\bv x,t')$ as follows:
\begin{eqnarray}
\fl \int_t^{t+\Delta t}dt'
\Pi^\alpha(\bv x,t')
&=-\frac{\partial}{\partial x^\beta}
\left[\int_t^{t+\Delta t}dt'
\int_0^{t'} ds e^{-(t'-s)/\tau}
\frac{1}{\gamma\rho(\bv x,s)}
\left(\frac{g^\alpha(\bv x,s)}{\tau}\right)
\left(\frac{g^\beta(\bv x,s)}{\tau}\right)
\right]\nonumber\\
\fl &=-\frac{\partial}{\partial x^\beta}
\left[\int_t^{t+\Delta t}dt'
\int_0^{t'} ds e^{-(t'-s)/\tau}
\frac{\Xi^\alpha(\bv x,s)\Xi^\beta(\bv x,s)}{\gamma\rho(\bv x,s)}
\right]+o(\Delta t).
\label{g:pre}
\end{eqnarray}
In the second step,
we have used the estimation $\Xi^\alpha(\bv x,t)\propto \tau^{-1/2}$
obtained from the following identity:
\begin{eqnarray}
\Xi^\alpha(\bv x,t_1)\Xi^\beta(\bv x,t_1)
=&2\gamma T\delta_{\alpha\beta}
\int_{0}^{t_1}ds
\frac{e^{-2(t_1-s)/\tau}}{\tau^2}[\rho(\bv x,s)]^2,
\label{identity:rhorho2}
\end{eqnarray}
which is obtained by substituting 
$t_1=t_2$ and $\bv x=\bv x'$ into 
equation (\ref{xixim1}) 
using equation (\ref{identity:rhorho}).

Substituting equation (\ref{identity:rhorho2}) into equation (\ref{g:pre}),
the integration of $\Pi$ is given as follows:
\begin{eqnarray}
\fl \int_t^{t+\Delta t}dt'\Pi^\alpha(\bv x,t')\nonumber\\
\fl =-\frac{\partial}{\partial x^\alpha}
\Big[
\int_t^{t+\Delta t}dt'
\int_0^{t'}dse^{-(t'-s)/\tau}\frac{1}{\gamma\rho(\bv x,s)}
\int_0^s ds_1\frac{e^{-2(s-s_1)/\tau}}{\tau^2}
2\gamma T[\rho(\bv x,s_1)]^2
\Big]\nonumber\\
\fl  + o(\Delta t).\label{start:pi}
\end{eqnarray}
Taking the limit $\tau\to 0$ in
equation (\ref{start:pi})
and by using the identity (\ref{formula:deltaintegral}),
we obtain 
\begin{eqnarray}
\lim_{\tau\to 0}
\int_t^{t+\Delta t}dt'\bv\Pi(\bv x,t')
=-T\bv\nabla\tilde\rho(\bv x,t)\Delta t+o(\Delta t).\label{result:Pi}
\end{eqnarray}
This evaluation shows that
the coarse graining of the 
advection term yields the diffusion term 
in the fluctuating hydrodynamics model of Brownian dynamics.

Finally, we obtain the change in density 
from time $t$ to $t+\Delta t$
in the limit of $\tau\to 0$
by substituting 
equations
(\ref{result:Upsilon}) and (\ref{result:Pi}) into equation (\ref{rho:base}).
The change in density to the order of $\Delta t$ 
is written as
\begin{eqnarray}
\fl \lim_{\tau \to 0}[\gamma\left(\rho(\bv x,t+\Delta t)-\rho(\bv x,t)\right)]
=&-\Delta t\bv\nabla\cdot \left[\tilde\rho(\bv x,t)\bv\nabla
\left[-\frac{\delta H[\tilde\rho]}{\delta\tilde\rho(\bv x,t)}\right]
-T\bv\nabla\tilde\rho(\bv x,t)\right]\nonumber\\
\fl &-\lim_{\tau\to 0}\bv\nabla\cdot\int_t^{t+\Delta t}dt' \bv\Xi(\bv x,t')
+o(\Delta t).\label{pre:Dean}
\end{eqnarray}
By multiplying both sides of equation (\ref{pre:Dean}) by $\Delta t^{-1}$
and taking the limit $\Delta t\rightarrow 0$, we obtain
\begin{eqnarray}
\fl \pder{\tilde\rho(\bv x,t)}{t}
=-\frac{1}{\gamma}
\bv\nabla\cdot\left[-\tilde\rho(\bv x,t)\nabla
\left[\frac{\delta H_V[\tilde\rho]}{\delta \tilde\rho(\bv x,t)}\right]
-T\nabla\tilde\rho(\bv x,t)
+\sqrt{\gamma T\tilde\rho(\bv x,t)}\bv\zeta(\bv x,t)
\right].\label{DeanKawasaki}
\end{eqnarray}
Here, we can rewrite the noise term as
\begin{eqnarray}
\lim_{\Delta t\to 0}
\left[\lim_{\tau\to 0}\frac{1}{\Delta t}
\int_t^{t+\Delta t} dt'\bv\Xi(\bv x,t')\right]
=\sqrt{\gamma T\tilde\rho(\bv x,t)}\bv\zeta(\bv x,t)
\label{noise:coincide}
\end{eqnarray}
because (\ref{XiXi}) shows that intensity 
in the left hand side of (\ref{noise:coincide})
coincides with that in the right hand side of (\ref{noise:coincide}).
This is the fluctuating hydrodynamic equation 
for the density in the 
overdamped limit \cite{Dean}.
Using a technique similar to that used in the underdamped case,
we can obtain the steady-state distribution function 
written as 
\begin{eqnarray}
P_{\rm eq}[\tilde\rho]
\propto \exp\left(
-\frac{H_V[\tilde\rho]}{T}
-\int d\bv x\tilde\rho(\bv x)(\log\tilde\rho(\bv x)-1)\right).
\label{steadystatedistribution:over}
\end{eqnarray}

From the derivation of the Kawasaki-Dean formula (\ref{DeanKawasaki}),
we have found that diffusion is caused by 
the advection due to a random force.
In contrast, in the case of liquids,
diffusion is caused by the liquid itself.
The physical origin of the diffusion term 
obtained by our model, therefore,
is different from that in a liquid system although the expressions of
in both the cases appear similar.

\section{Discussion}\label{concludingremarks}

The primary objective of this study is 
the derivation of the underdamped nonlinear 
fluctuating hydrodynamic equation (\ref{gdev}) along with 
equations (\ref{rhodev}), (\ref{spacetimenoise}),
(\ref{def:HV}), and (\ref{def:HK}).
The starting point is the underdamped Langevin
equations (\ref{xdev}) and (\ref{pdev}). 
It is a nontrivial fact that we obtain the closed stochastic evolution equation 
of the density field and momentum density fields
from the particle description model.
The exact derivation would have been obtained
if we had taken the continuous limit
using the discretization scheme discussed in
\ref{justification}.
The obtained evolution equation is reasonable because 
the Fokker-Planck equation obtained using our model 
agrees with that obtained using a classical liquid system, except 
for the form of the dissipative matrix \cite{Kawa94}.

The underdamped equation is unrealistic
when describing the experimental situation of Brownian particle system.
The overdamped model is more suitable 
as compared to the 
underdamped model, for a Brownian particle system.
Nevertheless, the underdamped model is 
useful for theoretical approximations 
such as the mode-coupling theory.
The underdamped model for the Brownian particles
can be a basis for the derivation of the
mode coupling equation \cite{Dawson}.
Recently, some researchers have developed systematic methods
for the derivation of the mode-coupling equation 
from the overdamped model for a Brownian particle system \cite{Kim}.
However, the derived equation is slightly 
different from the mode-coupling equation \cite{Kim}.
The difference might be eliminated if the mode-coupling equation is derived
from the underdamped model.

In addition, the underdamped model fits a liquid system.
The moment of liquid particles should be explicitly treated in order to study
the phenomena observed before momentum relaxation.
In a liquid system, however, the field description 
in the nonlinear and underdamped cases has not been intensively studied.
The present equations 
(\ref{Newton:field1}) and (\ref{Newton:field2}) for $\gamma=0$ 
can be applied in this case.
They are useful for microscopic studies of a liquid system.

There are similarities and differences between our equations
and the Euler equation.
A point $\bv x$ in
equations (\ref{Newton:field1}) and (\ref{Newton:field2})
includes not more than one particle. 
In contrast,
a point $\bv x$ in the Euler equation 
includes many particles such that the thermodynamic variables
are well defined.
Equations (\ref{Newton:field1}) and (\ref{Newton:field2})
have been derived exactly, except for the condition (\ref{deltaij}).
In addition, Euler equation is based on the local equilibrium  
assumption. In contrast,
equations (\ref{Newton:field1}) and (\ref{Newton:field2}) 
can be derived without such assumptions.  
Thus, equations (\ref{Newton:field1}) and (\ref{Newton:field2})
can be used to describe the liquid that is not in the  
local equilibrium state.

In Sec. \ref{toDean}, we
have derived the Kawasaki-Dean formula by
 coarse-graining our model with respect to time
under the condition (\ref{identity:rhorho}).
Similar coarse graining methods
for the Fokker-Planck equation for the derivation of the equation of 
the density 
and momentum density describing a liquid system have been devised \cite{Kawa94}.
The coarse-graining method described in Ref. \cite{Kawa94} 
does not require the condition (\ref{identity:rhorho}).
Therefore, 
the condition (\ref{identity:rhorho}) is not required 
if we carry our coarse graining for the Fokker-Planck equation
derived in Sec. \ref{deri:Fokker-Planck}.
This will be investigated in our future study.

We have found inconsistencies between
the steady-state probability distributions 
(\ref{steadystatedistribution:und}) and (\ref{steadystatedistribution:over}).
In Sec. \ref{deri:Fokker-Planck},
we have derived the steady-state probability distribution 
(\ref{steadystatedistribution:und})
for the underdamped model.
We have also obtained the steady-state probability distribution
for the overdamped model by using equation
(\ref{steadystatedistribution:over}).
The probability distribution
(\ref{steadystatedistribution:over})
is not obtained by integrating equation
(\ref{steadystatedistribution:und})
with respect to $[\bv g]$.
We guess that the inconsistencies 
might be related to 
the singularity of the delta function 
in equations (\ref{def:rho}) and (\ref{def:g}).
However, 
the relation between the inconsistencies and the singularity
has not determined thus far.
We will address  these inconsistencies in our future study.

\ack{
We are grateful to H. Furusawa, K. Miyazaki, S. Sasa, and Y. Hyuga
for their helpful discussions of this work.
This study was supported by the Grant-in-Aid for Scientific
Research on Priority Area and for Scientific Research (C)
from the Japanese Ministry of Education, Science, Sports and Culture.}

\begin{appendix}
\section{Justification of equations
(\ref{deltaij}), (\ref{Mdef2})
(\ref{Mdef3}) and (\ref{identity:rhorho})}
\label{justification}
\subsection{Discretization method and justification of (\ref{deltaij})}
In this study, we have often treated 
the delta function in the manner which is not 
mathematically well defined.
In this section, 
we give the correct interpretation of these treatments and representations.

First, we design a discretized cell $|\Delta \bv x|$
that has a finite size
is so small that different particles cannot occupy the same cell.
Such a situation can be considered if
the potential has a repulsive core within
a short length $r_c$.
We denote the position of the cell by $\bv I$ 
introduced in Sec. \ref{deri:Fokker-Planck}.
Because the cell size is sufficiently small,
the map from $i$ to $\bv I$ is an injective map.
Then, the following equality is satisfied:
\begin{eqnarray}
\frac{1}{|\Delta \bv x|}
\delta_{\lfloor \bv x_i/\Delta \bv x\rfloor,
\lfloor \bv x_j/\Delta \bv x \rfloor}
=\frac{1}{|\Delta \bv x|}
\delta_{\lfloor \bv x_i/\Delta \bv x\rfloor,
\lfloor \bv x_j/\Delta \bv x\rfloor}\delta_{i,j}
\label{disc:deltaij},
\end{eqnarray}
where $\lfloor a \rfloor$ is Gauss's notation
representing the maximum integer less than $a$
and 
$\delta_{
\lfloor \bv x_i/\Delta \bv x\rfloor,
\lfloor \bv x_j(t)/\Delta \bv x\rfloor}
\equiv\prod_{\alpha=1,2,3}\delta_{
\lfloor x_{i,\alpha}(t)/\Delta x^\alpha\rfloor,
\lfloor x_{j,\alpha}(t)/\Delta x^\alpha\rfloor}$.
When we take the continuous limit $|\Delta x|/r_c^3\to 0$,
equation (\ref{disc:deltaij}) converges to equation (\ref{deltaij}).

\subsection{Justification of 
equations (\ref{Mdef2}) and (\ref{Mdef3})}\label{justification1}
The evaluation of equation (\ref{Mdef2}) leads to its infinite form.
This infinite form is also justified by discretization.
First, we represent 
$\rho$ in equation (\ref{def:rho}),
$g^\alpha$ in equation (\ref{def:g}) and $M^{\alpha\beta}$ 
in equation (\ref{Mdef})
in the discretized form as follows: 
\begin{eqnarray}
\rho_{\bv I,t}&=&\sum_{i=1}^N
\frac{\delta_{\bv I,\lfloor \bv x_i(t)/\Delta \bv x\rfloor}}
{|\Delta \bv x|},\label{disc:rho}\\
g^\alpha_{\bv{I},t}
&=&\sum_{i=1}^N
p_i^\alpha(t)
\frac{\delta_{\bv I,\lfloor \bv x_i(t)/\Delta \bv x\rfloor}}
{|\Delta \bv x|}\label{disc:g},\\
M^{\alpha\beta}_{\bv{I},t}
&=&\sum_{i=1}^N\frac{p_i^\alpha(t)p_i^\beta(t)}{m}
\frac{\delta_{\bv I,\lfloor \bv x_i(t)/\Delta \bv x\rfloor}}
{|\Delta \bv x|}\label{disc:M1}.
\end{eqnarray}
Trivially, by taking the continuous limit mentioned above,
$\rho_{\bv I,t}$ and $\bv g_{\bv I,t}$ converges to $\rho(\bv x,t)$
and $\bv g(\bv x, t)$, respectively.

Using these discretized forms and
by dividing the cell position $\bv{I}$ into two cases, 
we will prove that
\begin{eqnarray}
M^{\alpha\beta}_{\bv{I},t}
=&\frac{g^\alpha_{\bv{I},t}g^\beta_{\bv{I},t}}{m\rho_{\bv{I},t}}.
\label{disc:Mdef3}
\end{eqnarray}
In the first case, consider $i$
such that $\bv{I}=\lfloor \bv x_i(t)/\Delta \bv x\rfloor$.
In the second case, $\bv{I}\ne\lfloor \bv x_i (t)/\Delta \bv x\rfloor$
at any value of $i$.
Equation (\ref{disc:Mdef3}) corresponds
to equation (\ref{Mdef3}) in the continuous limit.

In the case of
$\bv{I}=\lfloor \bv x_i(t)/\Delta \bv x\rfloor$, we can prove that the
left- and right-hand sides of 
equation (\ref{disc:Mdef3}) are  equivalent.
In this case, we can calculate the left-hand side of
equation (\ref{disc:Mdef3}) from 
equation (\ref{disc:M1}), so that
\begin{equation}
M^{\alpha\beta}_{\bv{I},t}
=\frac{p_i^\alpha(t)p_i^\beta(t)}{m|\Delta \bv x|}.\label{left}
\end{equation}
Since equations (\ref{disc:rho}) and (\ref{disc:g})
reduce to $\rho_{\bv I,t}=1/|\Delta \bv x|$
and $g^\alpha_{\bv{I},t}=p_i^\alpha(t)/| \Delta \bv x|$ respectively,
we have
\begin{equation}
\frac{g^\alpha_{\bv{I},t}g^\beta_{\bv{I},t}}{m\rho_{\bv{I},t}}
=\left(\frac{p_i^\alpha(t)}{|\Delta \bv x|}\right)
\left(\frac{p_i^\beta(t)}{|\Delta \bv x|}\right)
\left(\frac{|\Delta \bv x|}{m}\right)
=\frac{p_i^\alpha(t)p_i^\beta(t)}{m|\Delta \bv x|}.\label{right}
\end{equation}
Equations (\ref{left}) and (\ref{right}) are equivalent.

In the case of $\bv{I}\ne\lfloor \bv x_i(t)/\Delta \bv x\rfloor$ 
at any value of $i$, 
we prove equation (\ref{disc:Mdef3}) by defining the right-hand
side by zero.
From equation (\ref{disc:M1}) and
$\delta_{I^\alpha,\lfloor x_{i}^{\alpha}(t)/ \Delta \bv x\rfloor}=0$, 
we obtain
$M^{\alpha\beta}_{\bv{I},t}=0$.
The right-hand side of 
equation (\ref{disc:Mdef3}) is defined by zero
although it has an infinite form because $\rho_{\bv I,t}=0$.

\subsection{Justification of 
equation (\ref{identity:rhorho})}
The left-hand side of
equation (\ref{identity:rhorho})
is not well defined mathematically
because of the singularity.
This singularity is eliminated by the discretization of
equation (\ref{identity:rhorho}).
By using equation (\ref{disc:rho}), 
the product of $\rho_{\bv I,t}$ is easily calculated as
\begin{eqnarray}
\rho_{\bv I,t}\rho_{\bv I,t}
=&\frac{1}{|\Delta \bv x|^2}
\sum_{i,j=1}^N
\delta_{\bv I,\lfloor \bv x_i(t)/\Delta \bv x\rfloor}
\delta_{\lfloor \bv x_i(t)/\Delta \bv x\rfloor,\lfloor \bv x_j(t)/\Delta \bv x\rfloor}.
\label{rhorho:0}
\end{eqnarray}
Substituting equation (\ref{disc:deltaij}) into equation (\ref{rhorho:0})
and taking the summation with respect to $j$,
we obtain
\begin{eqnarray}
\rho_{\bv I,t}\rho_{\bv I,t}
=&\frac{1}{|\Delta \bv x|^2}
\sum_{i=1}^N
\delta_{\bv I,\lfloor \bv x_i(t)/\Delta \bv x\rfloor},\label{disc:rhorho}
\end{eqnarray}
In the continuous limit,
equation (\ref{disc:rhorho}) corresponds to equation (\ref{identity:rhorho}).

\end{appendix}

\end{document}